# From Efficiency to Sufficiency: The Cost-Aware Measurement Design Framework for Automated Essay Scoring

Yi Gui ygui@measinc.com

Measurement Inc.

## Abstract

Compute cost in automated essay scoring (AES) is usually treated as a technical efficiency issue rather than a measurement design concern. Yet deployment decisions require more than reporting latency, model size, or inference expense. In high-stakes settings, total burden includes validation, monitoring, adjudication, compliance, and governance; in instructional settings, pricing, licensing, and integration may dominate cost. This paper reframes AES as a cost-constrained measurement system and introduces the Cost-Aware Measurement Design (CAMD) Framework, a four-layer structure—metering, allocation, unit economics, and decision—for translating computational and workflow resource records into decision-relevant cost constructs. Direct cost is specified as monetary outlay plus attributable labor and time, and two criteria—design attribution and auditability—determine what the cost boundary excludes; burdens that satisfy neither, such as institutional capability and governance posture, are represented as constraints on which designs are admissible rather than priced. It defines cost per defensible score (CPDS) and argues that AES systems should be compared by quality-cost sufficiency: minimizing cost among defensible scoring designs that satisfy measurement requirements. Scoring volume conditions all four layers, so cost per defensible score is treated as a function of scale rather than as a single transferable number. CAMD provides a portable basis for representing, comparing, and evaluating cost across AES studies and deployment contexts.

# From Efficiency to Sufficiency: The Cost-Aware Measurement Design Framework for Automated Essay Scoring

Automated essay scoring (AES) research has long emphasized efficiency, including reduced human scoring burden, faster turnaround, and improved scalability. Yet the field still lacks a stable language for representing compute cost beyond proxy indicators such as speed, model size, or reduced labor. As a result, efficiency claims remain difficult to compare, transfer across settings, or interpret within measurement-oriented decision making (Dikli, 2006; Bennett, 2004; Rotou et al., 2020).

Current deployment conditions make this limitation increasingly consequential. In high-stakes programs, the scoring engine is only one component within a broader system of psychometric validation, security, and compliance, where cloud inference may represent a minor portion of total cost relative to adjudication, validity studies, auditability, and governance. In instructional and formative contexts, by contrast, cost structures often resemble SaaS models, where licensing, API pricing, and platform integration dominate (Educational Testing Service, 2021; Fang et al., 2025; Bai & Hui, 2024). Across both contexts, cost behaves as a system-level construct rather than a model-level metric.

This paper starts from a simple observation: the problem is not that AES ignores cost altogether, but that it lacks a framework for representing cost in a way that is portable, comparable, and decision-relevant. In cloud operations and machine learning systems, cost is routinely metered, allocated, and normalized into units that support decision making. By contrast, AES studies still often rely on isolated indicators that do not distinguish direct from shared burden, total from unit cost, or technically cheap from operationally sufficient (FinOps Foundation, n.d.-a, n.d.-c; OpenCost Specification, n.d.; Li et al., 2023). This paper supplies that framework. It introduces Cost-Aware Measurement Design (CAMD), which represents cost through four cumulative layers, defines cost per defensible score (CPDS) as its central unit, and argues that scoring designs should be compared under quality-cost sufficiency rather than by raw performance. The section that follows establishes the gap CAMD is built to address; the remainder of the paper develops the framework and its implications.

## The AES Gap: Efficiency Without Decision-Relevant Cost Representation

Existing AES literature has not failed to recognize efficiency, but has largely failed to represent it in a way that is portable, comparable, and decision-relevant across deployment contexts. Early AES reviews explicitly frame automated scoring as a response to the labor, time, and financial burden of hand scoring. Operational discussions of AES deployment emphasize continuous monitoring, validation, and quality control. Measurement theory has long treated cost, feasibility, and utility as legitimate dimensions of assessment design. Educational cost analysis offers established methods for identifying ingredients and estimating cost-effectiveness. Efficient NLP and Green AI research, meanwhile, have normalized the idea that computational burden should be reported rather than treated as invisible background. What remains missing is the integration of these strands into a portable compute-cost framework for modern AES (Dikli, 2006; Rotou et al., 2020; van der Vleuten, 1996; Schuwirth & van der Vleuten, 2011; Levin et al., 2018; Institute of Education Sciences, 2020; Schwartz et al., 2020; Tay et al., 2022). Table 1 summarizes these strands and the specific cost-representation gap each leaves unresolved.

**Table 1.**

*Representative strands relevant to AES cost analysis*

| Literature strand | Main contribution | What remains missing for AES |
|---|---|---|
| Early and general AES literature | Recognizes labor reduction, faster scoring, and operational scale as core motivations for automation. | Does not formalize compute cost as a structured construct or provide portable reporting rules for contemporary model deployments. |
| AES deployment guidance | Emphasizes validation, monitoring, fairness, and quality control in operational use. | Usually treats cost as a background consideration rather than a design variable with explicit boundaries and allocation rules. |
| Psychometric and assessment-utility theory | Shows that assessment quality must be considered alongside feasibility, cost, and use. | Does not yet specify how compute-related burden should be represented in transformer-based scoring systems. |
| Educational cost analysis | Provides ingredients-based costing and cost-effectiveness logic. | Has seen limited extension to modern AES deployments where cost depends on model serving, storage, orchestration, and reserved capacity. |
| Efficient NLP and Green AI | Normalizes reporting of latency, energy, memory, and efficiency trade-offs. | Usually evaluates burden at the model or system level without grounding comparisons in measurement sufficiency for assessment use. |

The gap can be stated more sharply. Across the relevant AES and adjacent strands, the dominant reporting vocabulary still centers on score agreement, latency, model size, and sometimes reduced human effort. Those metrics are useful, but they do not by themselves constitute a cost representation language. They rarely specify the boundary of cost, how shared or idle burden is treated, or which unit of output should anchor comparison (Dikli, 2006; Rupp et al., 2019; Rotou et al., 2020). Table 2 illustrates this reporting gap by contrasting common AES efficiency reporting with the elements of a structured cost representation.

**Table 2.**

*Stylized gap demonstration: common AES efficiency reporting versus structured cost representation*

| Often reported or discussed | Rarely made explicit | Why the omission matters |
|---|---|---|
| Model size, runtime, latency, throughput, or reduced human scoring time | Formal cost boundary | Without a boundary, it is unclear whether the claim concerns direct inference only or a broader operational workflow. |
| Aggregate performance and efficiency claims | Allocation rules for shared, idle, or overhead cost | Operationally meaningful comparison requires specifying what portion of infrastructure burden belongs to the scoring workload. |
| Technical metrics for a specific experiment | Portable unit economics such as cost per essay, cost per defensible score, or cost per acceptable configuration | Stakeholders need interpretable workload units rather than raw engineering signals alone. |
| Accuracy-oriented ranking of models | Decision rules linking cost to validity, reliability, fairness, and feasibility thresholds | Measurement design rarely chooses systems by score quality alone; sufficiency under intended use is the actual decision problem. |

Taken together, these patterns suggest that current AES efficiency reporting lacks three properties required for meaningful cost reasoning: (1) an explicit cost boundary, (2) a principled allocation of shared and indirect burden, and (3) a decision-relevant unit of analysis. Without these elements, efficiency claims remain difficult to interpret, compare, or operationalize in real assessment settings.

The gap is not that every AES paper ignores cost in the same way, but that the field lacks a shared representational grammar for compute burden—one this framework is designed to supply.

The limitation of prior approaches is therefore not simply omission, but representational mismatch: existing metrics describe computational properties, whereas AES requires cost constructs that operate at the level of defensible scoring outcomes.

## The Present Paper: Motivation and Contributions

Although conceptual in form, the present framework is motivated by recurring patterns in automated essay scoring research on representation compression, ensemble variability, and scoring stability. First, substantial reductions in representation dimensionality often yield minimal degradation in scoring performance. Second, variability induced by stochastic training factors (e.g., random seeds) can exceed the gains achieved through architectural modifications. Third, design choices that appear efficient at the model level can increase downstream operational burden, particularly through instability and review requirements.

While detailed results are outside the scope of this conceptual paper, these recurring patterns motivate the need for a framework that connects computational cost to defensible scoring outcomes.

These observations point to a structural limitation. In AES, cost cannot be treated as an unconstrained optimization objective. Unlike many engineering or commercial settings, scoring systems must first satisfy minimum requirements for validity, reliability, and fairness before cost considerations become meaningful. The relevant problem is therefore not how to minimize computational expense in isolation, but how to reason about cost within a constrained measurement space.

This paper proposes the Cost-Aware Measurement Design Framework, a measurement-informed framework for cost-aware AES that makes cost explicitly comparable across designs while anchoring cost evaluation in measurement requirements.

To make this distinction concrete, consider a program evaluating two AES configurations for operational use. The first relies on a lower-cost model with minimal per-inference expense but produces less stable scores, requiring additional monitoring and occasional human review. The second uses a more computationally expensive model that yields more consistent outputs, reducing the need for downstream intervention.

If cost is evaluated only at the level of model inference, the first configuration appears more efficient. However, once monitoring, review, and operational overhead are taken into account, the total cost required to produce a defensible score may differ substantially. This example illustrates a central issue: in AES, cost is often evaluated at the level of computation rather than at the level of usable scoring outcomes.

This paper makes three contributions. First, it reframes AES as a cost-constrained measurement system rather than treating cost as an external engineering consideration. Second, it introduces CAMD, a four-layer framework—metering, allocation, unit economics, and decision—that transforms computational and workflow resource records into decision-relevant cost constructs. Third, it defines cost per defensible score (CPDS) and formalizes quality-cost sufficiency as the appropriate basis for comparing AES designs. The original contribution of CAMD lies not in inventing new cost-accounting machinery but in translating mature tools from cloud cost management, carbon accounting, and machine learning systems research into a measurement-centered framework for AES.

## Defining the Cost Boundary

Because this paper presents an initial framework for cost-aware AES, it does not attempt to model every possible burden at once. Doing so would blur the analytic target and make the framework harder to operationalize. For that reason, this paper treats direct operational cost over a declared scoring period as its primary scope. This boundary includes design-attributable preparation needed to put a scoring configuration into use, such as human scoring of training data and qualification, together with resources required to run and quality-assure scoring under deployment or near-deployment conditions. Those resources include compute for inference, attributable storage, orchestration, monitoring, and reserved capacity reasonably assigned to the scoring workload (Levin et al., 2018; Institute of Education Sciences, 2020; FinOps Foundation, n.d.-a; OpenCost Specification, n.d.).

Direct operational cost in this sense has two components, and the framework requires both. The first is monetary outlay: amounts the program is billed or pays, including inference charges, licensing, storage, and contracted scoring services. The second is labor and time: the hours of human effort that a scoring design consumes, including adjudication and backreading, reader recruitment, training, and qualification, quality-assurance review, monitoring, and the technical labor of deployment and recalibration. The direct component is therefore written as

$$C_{\text{direct}} = C_{\text{monetary}} + \Sigma_r\, w_r\, t_r$$

where $t_r$ is the time that role r spends on the scoring workflow and $w_r$ is the fully loaded hourly cost of that role. Separating the two channels is not cosmetic, because they behave differently under scale and under design change. Monetary outlay typically tracks scored volume through metered usage, whereas labor is bounded by staffing and by the rate at which qualified readers can be recruited and trained. A design that shifts burden from the first channel to the second can therefore appear inexpensive in a blended total while being infeasible at the volume a program actually serves, which is precisely the kind of substitution the framework is meant to make visible.

One convention keeps the two channels from being counted twice. Each resource cost is counted once, and labor already embedded in a vendor's charge is not entered a second time through the labor channel.

When a program pays a vendor a per-response price, that price already contains the vendor's labor and belongs wholly in monetary outlay, and no $w_r t_r$ term is added for it; when a program performs the activity with its own personnel, the effort enters through the labor channel and no corresponding line appears in monetary outlay. Mixed arrangements are decomposed activity by activity rather than in aggregate, since a program may contract first-pass review while adjudicating internally. Reporting that does not state which channel carries each activity cannot be audited, because the same dollar can be defended as either.

The labor channel carries weaker measurement properties than the monetary one, and the framework should say so rather than conceal it. Hours and rates are not equally available: review, adjudication, and monitoring time can often be recovered from program records, whereas fully loaded wage figures are internal, vary across institutions, and are frequently unavailable to researchers reporting on a deployment. Labor should therefore be reported twice. First, in natural units—reviewer hours, full-time-equivalent weeks, and review or adjudication rates per thousand responses—which are observable and transferable across institutions that pay differently. Second, in monetized form under rates that are declared as assumptions rather than measurements and varied across low, base, and high scenarios. The distinction governs what a study is entitled to claim: a conclusion that holds across the plausible range of rates is a conclusion about the scoring design, whereas one that reverses within that range is a conclusion about the rate and should be reported as such.

At minimum, a decision-useful AES cost boundary should be broad enough to include model serving, prompt and context burden, batchability, cacheability, storage and audit retention, human review volume attributable to machine uncertainty, and validation or monitoring overhead. That broader ingredients logic is important because the model call itself is often only one ingredient, and not always the most expensive one (Educational Testing Service, 2021; Rotou et al., 2020; OpenCost Specification, n.d.).

This boundary is intentionally narrower than the universe of all AI-related costs. Training compute may matter, especially when models must be re-trained or substantially fine-tuned; the human scoring of training data, by contrast, is treated below as part of the cost function rather than as a secondary extension. Carbon and energy reporting matter for sustainability and, in some cases, institutional decision making. Embodied infrastructure burden matters for full life-cycle accounting. But those burdens are best treated as secondary extensions in a first conceptual paper for AES. Their main role here is analogical and methodological: they show that mature fields already know how to reason about boundaries, indirect burdens, amortization, and shared infrastructure. They are not this paper's primary empirical target (Schwartz et al., 2020; Patterson et al., 2021; Lannelongue et al., 2021; CarbonTracker, n.d.; Google Cloud, n.d.).

Two criteria govern what falls outside the boundary, and stating them is more useful than enumerating exclusions. The first is design attribution. CAMD exists to compare alternative scoring designs for a given program, use, and operating scale, so the quantities that carry decision-relevant information are those that differ across the designs under comparison. A burden that is identical under every candidate design cancels in the comparison when the designs under comparison deliver the same number of defensible scores: it raises the level of a program's cost without changing which design is least costly among the sufficient ones. The condition is not decorative. Because CPDS divides by S(d, N), a common dollar amount enters as a larger per-score charge for a design with a lower defensibility yield, so when S differs across designs a common fixed burden can reorder them and must then be carried explicitly. The working rule follows directly: the more the candidate designs differ in the share of responses that reach a usable score, the less a common burden can be set aside. The second is auditability. A quantity enters the cost function only if a program could in principle produce a record for it—an invoice, a contract, a schedule, a time log, a routing rate—so that a reported CPDS can be recomputed and contested rather than merely asserted.

Applied jointly, these criteria sort the broader burdens that operational discussion raises into three positions rather than into inclusion or dismissal. Burdens that are both design-attributable and auditable enter the cost function, and this is where reader recruitment, training, and qualification, backreading and quality assurance, monitoring, and the technical labor of deployment belong. They belong there even when they are described as organizational rather than computational, because a design that routes a fifth of its responses to human readers requires a different scoring workforce than one that routes a twentieth, and that difference is both attributable to the design and recoverable from program records. Burdens that are real but satisfy neither criterion—institutional reputation, partnership and contracting posture, the quality of management policy, and accumulated institutional expertise—are not priced here, and they are not thereby treated as immaterial. Where they bear on a scoring decision, they bear on it by determining which designs a program is able to operate at all, so the framework locates them in the sufficiency constraint of the decision layer rather than in the numerator of CPDS. Burdens that are plausibly auditable but outside the present scope, including training compute and carbon, remain secondary extensions as described above.

Applied to the model-development side, the same two criteria cut more finely than a single training-cost line. Human scoring of training and calibration responses is design-attributable, because the number of responses a program has scored is a choice the program makes rather than a fact about the model, and it is auditable, because it leaves records: contracts and invoices where the scoring is outsourced, time and rate records where it is performed in house. It therefore enters the cost function. A qualification study run for a particular configuration is attributable and recorded on the same reasoning, and so is scheduled post-deployment recalibration, which enters as a recurring item rather than as a one-time charge because the design determines how often it recurs. Training compute divides along the same axis rather than along the line between human and machine effort. Compute for training or fine-tuning undertaken for a particular candidate design is in principle both attributable and auditable, since cloud billing and run records can support it. What resists attribution is development that precedes the comparison: pre-training of a base model, and equally a scored corpus or tuning run acquired earlier and reused across programs and uses, are shared historical investments whose allocation to a single scoring design is not recoverable from program records without a separately stated allocation rule. Design-specific training compute is therefore measurable in principle, but it is not quantified in the present illustration, and that is a choice of scope rather than a claim about what can be measured. Historical and shared development cost carries the further requirement of an explicit allocation rule, which this first statement of the framework does not attempt to settle.

Constraining the boundary in this way has two advantages. First, it improves tractability. Researchers can begin with a limited ingredients-style model of operational burden even when they do not have access to full billing exports or machine-level telemetry. Second, it improves portability. Institutions may differ widely in pricing and deployment scale, but they can still share a common schema for what cost components should be counted and how those components should be combined (FOCUS Specification, n.d.; Institute of Education Sciences, 2020).

## External Literatures and Emerging AES Evidence

CAMD draws on several external literatures, but their role in it is deliberately bounded. Table 3 translates these mature methods into AES design language. FinOps contributes the language of unit economics and the distinction between resource-side metrics and workload-side metrics. FOCUS contributes the idea that portability can exist at the level of schema and accounting semantics even when prices differ across institutions. OpenCost contributes an explicit logic for handling shared, idle, and overhead burden in a way that is more realistic than narrow inference accounting. Green AI and carbon-accounting frameworks contribute boundary discipline and the norm that computational burden should be reportable rather than invisible.

Machine learning systems research contributes the crucial insight that cost belongs in the design objective itself, often through frontier-based or constrained comparisons (FinOps Foundation, n.d.-a, n.d.-b, n.d.-c; FOCUS Specification, n.d.; OpenCost Specification, n.d.; Schwartz et al., 2020; Lannelongue et al., 2021; Li et al., 2023).

**Table 3.**

*Translating mature external methods into AES design language*

| External literature | Transferable concept | AES translation |
|---|---|---|
| FinOps and FinOps for AI | Unit economics; resource-side versus service-side metrics | Distinguish cost per GPU-hour from cost per essay or cost per acceptable-scoring configuration. |
| FOCUS | Portable schema and common cost semantics | Standardize cost components and formulas across AES studies even when dollar values differ by institution. |
| OpenCost | Allocation of direct, shared, idle, and overhead burden | Represent not only inference cost but also reserved capacity, monitoring, storage, and shared operational support. |
| Carbon-accounting frameworks | Boundary discipline; treatment of indirect and amortized burden | Clarify what is included in operational AES cost and what is treated as a secondary extension. |
| ML systems research | Cost-performance frontiers; constrained optimization | Compare AES architectures under quality thresholds rather than only by absolute score quality or isolated speed claims. |

CAMD translates these existing tools into a measurement-centered framework for AES.

Emerging AES-specific evidence, although still thinner than the broader external literatures, points in the same direction. Direct cost analyses in writing-screening contexts suggest that automation becomes economically stronger as assessment volume and frequency increase. Recent work on incremental learning and compact multi-task architectures suggests that retraining cadence, storage, and maintenance debt are themselves meaningful cost drivers. Comparisons between traditional AES and LLM-based scoring also suggest that higher inference cost does not automatically buy better scoring performance, even when it may buy better explanations or richer feedback. Together, these patterns reinforce the same conclusion: AES requires a system-level cost language, not just a faster model (Matta et al., 2022; Bai & Hui, 2024; Fang et al., 2025).

## The CAMD Framework

CAMD has four layers: metering, allocation, unit economics, and decision. The layers are cumulative. Cost must first be observed, then assigned, then normalized, and only then used to support design decisions. CAMD is measurement-informed because the decision layer is explicitly anchored in validity, reliability, fairness, feasibility, and utility. Its purpose is to convert computational and workflow resource records into design-relevant constructs that can travel across real AES deployment settings. Scoring volume, denoted N, runs through all four layers rather than forming a layer of its own: it is metered as an observable, governs how fixed and variable components are allocated, conditions the unit in which cost is expressed, and shifts the thresholds against which sufficiency is judged.

### Metering

The metering layer records the observables that make cost analysis possible. The first of these is the scoring volume N, the number of essays scored in the run being metered, together with the deployment regime under which that run took place: workstation CPU, a single accelerator, a managed cluster, or a third-party API. N and regime are not incidental metadata. Every resource-side quantity below is interpretable only relative to them, because the same algorithm yields very different totals at N = 100 and at N = 1,000,000, and because the regime itself changes discontinuously as N grows. In a minimal AES setting, the resource-side observables would likely include GPU-hours or CPU-hours, memory use, storage footprint, throughput, latency, queue or wait time, request volume, and configuration-specific information such as model size, prompt length, embedding dimensionality, or compression setting. Because direct cost is defined to include labor and time, the same layer must also record the workflow observables that the labor channel requires: the realized share of responses routed to human review, hours by role and task for review, adjudication, and monitoring, the volume of adjudication beyond first review, and the effort invested in reader training and qualification. Specifying these fields is not itself a separate research design, since the quantities they name arise from the scoring workflow. Obtaining their values, however, is not always a matter of logging alone: routing logs and realized review volumes are often recoverable from routine operational records, while task-level hours may require new time measurement, and fully loaded labor rates come from payroll or contracting records rather than from the scoring workflow, in either case information that may not be available to an external researcher. Where such records are unavailable and a value must be assumed instead, the assumption should be reported as one: a cost claim built on assumed hours and a cost claim built on recorded hours do not carry the same evidential weight, and a framework that does not distinguish them invites the reader to treat an illustration as an estimate. AES developers need not meter everything. They should instead identify a stable core of observables that can be reused across studies and translated into comparable cost ingredients (MLCommons, n.d.; Lannelongue et al., 2021; Patterson et al., 2021).

For example, in an AES system using an API-based scoring model, metering may include token usage per essay, request latency, and batch throughput, alongside infrastructure-level signals such as CPU utilization and memory allocation. On the workflow side it may include the proportion of responses flagged for human review, the time readers spend per flagged response, and the number of cases escalated to adjudication. Together these quantities determine the measurable cost footprint of a scoring configuration.

**Allocation**

Metering alone does not produce interpretable cost comparison. AES workloads may run in shared environments, reserve capacity beyond their realized use, or depend on orchestration and monitoring systems that do not show up in a narrow inference window. The allocation layer therefore specifies which components count as direct burden, which are shared, which are idle or reserved, and how those components are attributed. Allocation must also separate cost that is fixed with respect to N from cost that varies with it. Environment preparation, deployment validation studies, reserved capacity, and scheduled recalibration are fixed with respect to N within a declared time window or scoring cycle and are then spread across however many essays are scored in it; per-inference compute, expected human review, and per-essay storage grow with N. Fixed in this sense means invariant to volume rather than incurred only once: reserved capacity and scheduled recalibration recur on their own schedules, so the window over which a fixed component is amortized has to be stated alongside N. Reporting the two separately is what makes a cost claim portable, because a reader can recover the per-essay figure for their own volume by dividing the fixed component by their N and adding the variable component, provided the pricing and workflow assumptions are stated. What must be held constant for a comparison to be meaningful is the scoring task, the use, the volume, the defensibility threshold, and the cost boundary; the deployment regime need not match, since the choice between local serving and a third-party API is itself one of the design alternatives the framework exists to compare. A single blended figure cannot be transferred in this way, since it silently encodes the volume at which it was produced. This step is essential if cost claims are meant to speak beyond a single benchmark run (OpenCost Specification, n.d.; Li et al., 2023).

For instance, in a shared cloud environment, GPU capacity may be reserved to meet latency requirements even when utilization is low. Similarly, monitoring pipelines, logging systems, and audit storage may incur cost without being directly attributable to individual scoring calls. Allocation rules therefore determine how these indirect and shared components are assigned to the AES workload.

**Unit Economics and Cost per Defensible Score (CPDS)**

Once cost components have been observed and allocated, they must be normalized into units that matter for operational and assessment decisions. In AES, candidate units include cost per essay, cost per 1,000 essays, cost per batch, cost per scoring day, or cost per configuration that reaches an agreed level of measurement quality. This layer matters because raw technical signals rarely communicate well with program administrators, procurement stakeholders, or assessment designers. Unit economics turns engineering burden into operationally interpretable burden (FinOps Foundation, n.d.-c; Levin et al., 2018). Every unit in this list is conditional on N. Because fixed cost is spread over the scored volume while operational burden such as monitoring, adjudication, and governance need not grow proportionally with it, cost per essay is not a constant but a function of N, commonly falling as fixed cost is absorbed and not necessarily monotonic thereafter. CPDS should therefore be reported as CPDS(N), accompanied by the volume and deployment regime at which it was obtained. A single CPDS value is evidence about one point on that curve; it does not license comparison with a value obtained at a different scale. This has a practical consequence for reporting. Where feasible, cost should be reported at more than one scoring volume, for example the full corpus and a subsample an order of magnitude smaller. Within a single deployment regime, two volumes allow a reader to recover the implied fixed and variable components, since the difference in total cost between the two runs is attributable to the variable component alone, and to check that recovery against the decomposition the authors report. This is a consistency check rather than an estimate. It assumes that cost is approximately linear in N across the interval the two runs span, an assumption that fails wherever the deployment regime itself changes, so disagreement between the reported and the recovered decomposition is itself informative: it indicates either an error in the decomposition or a regime shift inside the reported range. Formally, CPDS is the expected total cost of a scoring run divided by the expected number of scores that run delivers for use:

$$\mathrm{CPDS}(d, N) = \frac{\mathrm{E}[C_{\mathrm{total}}(d, N)]}{\mathrm{E}[S(d, N)]}$$

Here the numerator is the expected total cost of scoring N responses under configuration d, and the denominator, S(d, N), is the expected number of scores that configuration delivers for use. S equals N when every response completes the workflow, and is smaller when responses are routed out as unscoreable, left unresolved in adjudication, or withdrawn on appeal; the cost of those responses is incurred without yielding a usable score, and the ratio makes that visible. The configuration d is restricted throughout to those that satisfy the sufficiency thresholds set out in the decision layer below, so the quality conditions are imposed on the configuration rather than on any individual score.

Throughout, a defensible score is shorthand for a score produced under a configuration that has been shown, for a declared interpretation and use, to satisfy predeclared thresholds on validity-relevant evidence, reliability, fairness, and operational adequacy. The wording is deliberate. Validity in current measurement theory is a property of the interpretation and use of scores rather than of scores or instruments themselves, so a score cannot be valid in isolation (Messick, 1989; Kane, 2013; AERA, APA, & NCME, 2014). CPDS accordingly prices the conditions under which a score can be defended for a stated use; it does not treat defensibility as a property the score carries on its own.

CPDS challenges the assumption that cost per inference is the relevant economic unit. In operational AES settings, a model output is not yet a defensible score: it may still trigger human adjudication, monitoring, audit logging, error correction, appeal handling, or subgroup review (Rotou et al., 2020; Educational Testing Service, 2021; Matta et al., 2022).

Cost per inference is therefore not the same as cost per defensible score. Lower raw inference cost does not necessarily imply lower effective cost. To make this distinction more concrete, consider a simplified illustrative decomposition of two AES configurations that differ in both inference cost and downstream operational burden. Configuration A uses a model with lower per-inference cost but produces less stable scores, resulting in a higher rate of human review. Configuration B uses a model with higher inference cost but yields more consistent outputs, reducing the need for downstream intervention.

A stylized but operationally interpretable comparison is shown in Table 4. Although the numeric values are illustrative rather than tied to a specific deployment, they reflect realistic cost components observed in operational AES settings. This illustration compares the two configurations at the same scoring volume and under a common deployment regime to isolate the effects of inference and human review costs. The framework also permits comparisons across deployment regimes when the scoring task, intended use, sufficiency requirements, volume, and cost boundary are aligned.

**Table 4.**

*Stylized cost-per-score illustration: lower inference cost does not necessarily yield lower effective cost*

| Component | Configuration A (lower-cost inference) | Configuration B (higher-cost inference) |
|---|---|---|
| Inference cost per essay | $0.01 | $0.05 |
| Human review rate | 20% | 5% |
| Cost per human-reviewed essay | $0.50 | $0.50 |
| Expected review cost | $0.10 | $0.025 |
| **Selected-component cost per score** | **$0.11** | **$0.075** |

Note. The figures are a selected-component cost per score rather than a CPDS estimate: they price inference and first-pass human review only, and omit monitoring, validation, adjudication beyond first review, audit retention, and the human scoring of training data and qualification costs, all of which fall inside the declared cost boundary and are treated under Defining the Cost Boundary above. Values assume the same scoring volume N, the same deployment regime, and S(d, N) = N for both configurations. The cost per human-reviewed essay is stated as a contracted per-response price and therefore already includes the reviewer's labor, so no separate labor term is added for it. For orientation, at an assumed fully loaded labor rate of $50 per hour, $0.50 is equivalent to thirty-six seconds of labor and six minutes of labor is equivalent to roughly $5.00. The conversion states the pricing logic only: it does not identify the time or service content embedded in any particular vendor's price. A design requiring full second scoring rather than a brief confirmatory check would price at a different order of magnitude, and the comparison below would change accordingly.

Although the numerical values in Table 4 are illustrative, comparable cost structures have been documented in operational AES systems, where human review, monitoring, and governance processes account for a substantial portion of total scoring cost, particularly in large-scale assessment programs where human adjudication and quality control constitute nontrivial operational costs.

This example highlights a key implication of the CPDS concept. Within the variable costs listed, and therefore at any scoring volume, Configuration A carries the higher cost per score on the components listed despite its lower per-inference cost, because downstream review dominates the difference; Configuration B carries the lower one even though its raw inference cost is higher. The ranking is a ranking of those listed variable costs, and what follows shows what becomes of it once a fixed outlay that differs across the two configurations is restored.

Separating the training-side items exposes a structure that the fixed-and-variable distinction otherwise conceals: the two categories need not be independent. A program that invests more in human-scored training data may face a lower variable rate downstream, in which case the review rate in the cost function is not an exogenous property of a model but a quantity the design partly buys. The illustration that follows assumes such a relationship rather than establishing one: Configuration B is assumed both to require the larger investment in human-scored training data and to carry the lower human-review rate. Whether, and under what conditions, additional training data reduces review requirements is an empirical question this paper does not answer. What the example shows is the consequence for cost comparison when a fixed outlay and a variable rate move together in this way.

Table 5 extends the same illustration by restoring that outlay. Once it is carried, the ranking depends on volume rather than on model quality alone: under the illustrative assumptions, Configuration B's higher fixed outlay is offset by its lower variable cost only above a crossover volume, and below that volume the configuration with the higher routing rate is the cheaper one. The point generalizes beyond these figures. Whenever a fixed outlay purchases a reduction in a variable rate, a CPDS comparison is interpretable only at a declared N, and a comparison reported without its volume is not a comparison of designs but of operating scales.

Table 5.

Volume dependence once training-data outlay is carried: the ranking in Table 4 holds only above a crossover scoring volume

| Component | Configuration A | Configuration B |
|---|---|---|
| Human-scored training responses | 2,000 | 8,000 |
| Cost per human-scored response | $3.00 | $3.00 |
| Training-data outlay (fixed) | $6,000 | $24,000 |
| Variable cost per essay (from Table 4) | $0.110 | $0.075 |
| **Selected-component cost per score at N = 100,000** | **$0.170** | **$0.315** |
| **Selected-component cost per score at N = 1,000,000** | **$0.116** | **$0.099** |

Note. Values add the training-data outlay to the components listed in Table 4 rather than replacing them, and they remain a selected-component cost per score rather than an estimate of CPDS or of total operational cost. Variable cost per essay is taken from Table 4 (inference plus expected first-pass review). Training-data outlay assumes a contracted price of $3.00 per human-scored response, covering two independent ratings and adjudication, and assumes a single scoring cycle with no reuse of the scored corpus across cycles or programs. Under these figures the two configurations are equal at a volume of approximately 514,000 responses; Configuration A is cheaper below it and Configuration B above it. Where the scored corpus is reused across cycles or programs, the crossover depends on how the outlay is allocated across those uses and on whether N denotes a single cycle or cumulative volume; stating that allocation rule is itself a design decision rather than an accounting detail.

Two conditions limit how far this comparison travels, and both were raised in operational discussion of the framework. First, the routing rate is treated here as a design variable, but some programs operate under contracts or regulations that constrain the share of responses receiving human review, and such a constraint binds differently depending on whether it fixes that share or sets a floor beneath it. Where the share is genuinely fixed, a more stable configuration does not reduce it, and the particular saving this comparison attributes to Configuration B, namely fewer responses reviewed, does not arise. The gain appears instead as lower disagreement within a fixed review volume, which may still reduce cost through other channels, including time per review, the volume of disputes, and adjudication beyond first review;

whether it does is a separate question that this illustration does not price. What a framework should not do is count the gain as reduced review volume, which would mislead exactly the programs least able to realize it in that form. Second, the comparison prices a single scoring cycle, and training evidence is among the most reusable assets a program holds.

The purpose of this illustration is not to estimate absolute cost, but to demonstrate how different design choices redistribute cost across components. In AES, cost is therefore not a property of the model alone, but of the full scoring process required to produce a defensible outcome.

A cheaper system may generate higher downstream burden if its outputs are uncertain, unstable, or non-defensible, increasing adjudication, monitoring, or governance cost.

In this sense, CPDS shifts the unit of cost from computation to defensibility. Rather than asking how much it costs to generate a score, the relevant question becomes how much it costs to produce a score that can be defended under validity, reliability, fairness, and operational constraints.

**Decision**

The final layer is where compute cost becomes explicitly psychometric. AES systems are not chosen only because one model has a slightly better score correlation or a slightly lower latency. They are chosen because they are adequate for an intended use under realistic constraints. That means cost comparison should be conducted under sufficiency thresholds rather than under unrestricted performance ranking (van der Vleuten, 1996; Schuwirth & van der Vleuten, 2011).

$$D^*(N) = \{d \in D: V(d) \geq v^*, R(d) \geq r^*, F(d) \geq f^*, O(d, N) \geq o^*\}$$

Then choose $d^*(N) = \underset{d \in D^*(N)}{arg\,min}\, CPDS(d, N)$.

Here, V, R, and F represent validity-relevant, reliability-related, and fairness-related evidence; O represents operational sufficiency, including timeliness and feasibility; and CPDS(d,N) is defined only after those thresholds and the operating scale have been declared. The exact indicators can vary by use case. Sufficiency is itself scale-dependent, and the operational threshold in particular should be read as a function of N. A design that routes every essay to a human reader may satisfy operational sufficiency at N = 50 and fail it outright at N = 500,000, where staffing and turnaround make it infeasible; conversely, a configuration provisioned for continuous high-volume scoring is wasteful at small N without being any more defensible. The framework therefore does not identify a best AES system. It identifies the least costly sufficient design at a declared operating scale. The conceptual point is that cost should discriminate among sufficient options, not excuse an insufficient one.

In plain terms, this means that cost should only be used to differentiate among scoring designs that are already adequate for the intended use, rather than to justify a design that fails to meet basic measurement requirements.

The sufficiency constraint is also where organizational conditions enter the framework. Staffing depth, procurement and contracting capacity, governance and audit readiness, and institutional experience with automated scoring rarely have defensible dollar values in a given program, but they determine which designs belong in the admissible set over which CPDS is minimized. A configuration that presumes continuous subgroup monitoring is not available to a program with no one to perform it, and a configuration that presumes rapid contracting for additional readers is not available to a program whose procurement cycle is measured in quarters. Representing these conditions as constraints on admissibility keeps them inside the analysis without requiring the framework to assert prices it cannot defend, and it preserves the asymmetry on which the decision layer rests: a constraint can disqualify a design, whereas cost can only discriminate among the designs that remain.

This formulation is also compatible with generalizability-theory-based design optimization. In such settings, reliability or dependability targets can be treated as constraints, while alternative scoring configurations—for example, number of models, ensemble size, rescoring passes, or human-review triggers—imply different cost profiles. This enables cost-aware design studies in which the objective is not only to achieve sufficient measurement quality, but to do so at minimal operational cost (Parkes, 2000; Jiang et al., 2021). The parallel runs deeper than shared vocabulary. A generalizability study estimates variance components that are defined relative to a declared universe of admissible observations, and a decision study then projects reliability at whatever number of raters, occasions, or tasks a program can afford, provided the projected design stays within that universe; the coefficient obtained from any one design does not transfer, but the components that generate it do (Brennan, 2001). Cost behaves the same way. The fixed and variable components recovered from runs at two or more volumes support projection to other volumes under comparable hardware, pricing, and workflow, whereas CPDS at a single N does not transfer even that far. Reporting cost components rather than a single blended figure is therefore the same methodological commitment that leads measurement researchers to report variance components rather than a coefficient obtained from one sample.

CPDS reframes cost from a property of computation to a property of defensibility.

## Measurement Principles Underlying the CAMD Framework

In the CAMD Framework, cost is embedded in the logic of assessment design rather than appended as an engineering afterthought. The CAMD Framework focuses on five relevant measurement dimensions.

1. **Validity.** Cost does not override intended score interpretation and use. A cheaper system is not preferable if the evidence base for intended inferences becomes weaker.
2. **Reliability.** A lower-cost design is attractive only if it maintains acceptable consistency, agreement, or generalizability for the scoring purpose.
3. **Fairness.** Cost-saving choices should be evaluated for their impact on subgroup performance, bias, and differential degradation across populations.
4. **Feasibility.** Time-to-score, staffing realities, monitoring burden, and infrastructure constraints are part of the operational adequacy of a design. Organizational capacity belongs to this dimension as well, since staffing depth, procurement cycles, and governance readiness determine which designs a program can operate at all, independently of what those designs cost.
5. **Utility.** The desirable design is rarely the absolute best on one dimension. It is the one that best supports intended use under realistic trade-offs.

This psychometric anchoring changes the interpretation of cost. Cost is not a separate optimization problem sitting beside measurement quality. It is one of the constraints and design dimensions through which assessment systems become more or less defensible. In that sense, cost-aware AES is a natural extension of long-standing assessment utility arguments, not a foreign import from cloud engineering (van der Vleuten, 1996; Parkes, 2000; Schuwirth & van der Vleuten, 2011).

Compute cost becomes psychometric when it changes what scoring design is defensible.

## Research Agenda for Cost-Aware Measurement Design

Beyond reporting reform, the proposed framework opens a broader research agenda centered on cost-aware measurement design. That agenda is not limited to documenting cost; it concerns how cost interacts with quality thresholds, subgroup fairness, decision risk, and deployment context. Priority directions include the following.

**Empirical CPDS estimation across AES architectures.** Studies can compare traditional AES, transformer-based AES, and LLM-based scoring not only by agreement or validity evidence, but by the expected cost required to produce a defensible score.

**Cost-aware G-studies and D-studies.** Reliability or dependability targets can be treated as constraints while alternative scoring designs imply different cost profiles.

**Cost x fairness x decision-risk analysis.** Future work can examine whether apparently cheaper systems shift burden onto subgroup disparities, appeals, or human override processes.

**Cost-aware ensemble and hybrid-design selection.** The relevant question is not whether one scorer is globally best, but whether a hybrid, ensemble, or fallback design reduces CPDS in a given deployment context.

**Deployment-specific cost modeling.** High-stakes assessment, K-12 screening, classroom practice, formative revision, and tutoring contexts should not be assumed to share the same cost structure or the same defensibility thresholds.

**Cost reporting protocols and their validation.** A reporting protocol specifies the fields a study must disclose rather than the corpus it must run: scoring volume, deployment regime, batch configuration, the boundary of what is counted, fixed and variable components reported separately, human review rate, and the thresholds under which CPDS was computed. Work is needed both to fix that field set and to test it, by having independent programs describe the same system under the protocol and examining whether their cost claims become mutually recomputable.

**Public cost benchmarks, and what must precede them.** A shared corpus and reference deployment configuration would allow the third-party verification that self-report cannot provide. Existing research corpora can supply standardized essay inputs, but they do not by themselves standardize scoring volume, deployment regime, or the downstream adjudication, monitoring, and governance processes that dominate operational cost, so the components this framework treats as dominant would have to be simulated rather than measured. Because a corpus can be subsampled but not extended, the largest volume at which cost can be observed is bounded by corpus size, and publicly available AES corpora are small relative to operational deployment volumes, which is precisely the range in which operational burden becomes dominant. Protocol development would therefore need to come first, since a benchmark standardizes what is run rather than what is reported, and a common corpus without an agreed field set still yields incommensurable numbers.

Constraint-side evidence on organizational feasibility. The conditions this framework places in the sufficiency constraint rather than in the cost function—staffing capacity, procurement and contracting, governance readiness, and institutional experience with automated scoring—are empirically open. Characterizing them calls for an interview or survey design with its own sampling frame and human-subjects review rather than an extension of cost telemetry, and its product would be a better specification of which designs are admissible in which programs, not an additional term in CPDS.

The relationship between training volume and routing rate. The dependence assumed in the discussion of Table 5 is not established in this paper, and establishing it is a question of its own: whether, and under what conditions, additional human-scored training data reduces the share of responses requiring human review for a given model family, prompt type, and defensibility threshold, and where it does, how steep that function is. Estimating it is what would turn the volume-dependent comparison in Table 5 from an illustration into a design tool.

In the near term, the framework also supports immediate reporting reform. AES studies can move beyond isolated efficiency statements by reporting a minimum set of direct operational ingredients, an explicit cost boundary, and at least one interpretable unit-economic metric. Each of these should be accompanied by the scoring volume and deployment regime at which it was obtained, without which none of them can be recomputed by a reader operating at a different scale. By standardizing components and formulas rather than prices, AES research can support cumulative comparison across institutions that differ in hardware, procurement structure, and deployment volume.

Once the direct operational layer is established, the framework can be extended. Later work may incorporate broader human labor that is not attributable to a specific scoring configuration, training compute and retraining of the underlying model, carbon reporting, or amortized infrastructure cost. But those extensions should be built on top of a stable core, not mixed into the first statement of the framework in a way that makes the object of analysis indistinct.

This framework is subject to several limitations. First, it prices the model-development items that program records can support—human scoring of training and calibration responses, qualification studies, and scheduled recalibration—but leaves training compute, carbon, and embodied infrastructure outside the cost function, in the case of design-specific compute for reasons of scope rather than principle, and any of these may be material in a particular deployment context.

Second, CPDS depends on the specification of validity, reliability, and fairness thresholds, which may vary across programs and require further standardization.

Third, the framework is most directly applicable to medium- to large-scale AES systems, and its applicability to small-scale or low-resource settings may require adaptation.

Fourth, the reporting requirements implied by the framework are not themselves costless, and the framework should be willing to price them. Asking authors to disclose volume, regime, and a fixed-variable decomposition imposes a relatively low incremental reporting burden where studies already collect run-level telemetry, because it describes a run that would have been performed in any case. Mandating a common benchmark would impose a positive and recurring one, since the benchmark run is additional to the work a program actually needs to do and must be repeated as hardware and pricing change. That burden would not fall evenly. A fixed compliance cost is heaviest for those spreading it over the smallest volume, which is the same arithmetic the framework applies to scoring cost, now applied to the cost of complying with the framework itself.

Fifth, the two criteria that bound the cost function are themselves conditional. Design attribution is defined relative to a set of candidate designs within one program, so a burden that cancels in that comparison, such as institutional capability, would not cancel in a comparison across organizations, and a cross-organizational use of CPDS would require the common-mode components this paper sets aside. Auditability is a property of current record-keeping rather than of the burden itself, so quantities that cannot be priced defensibly today may become priceable as programs instrument their scoring workflows more fully. These limitations point to important directions for future work.

## Conclusion

The central claim of this paper is not that AES has overlooked efficiency. It is that AES still lacks a structured and psychometrically anchored language for discussing cost. That is why the field continues to oscillate between broad efficiency narratives and narrow engineering proxies. The CAMD Framework is offered as a way to address these limitations, making a needed conceptual advance before another round of empirical comparison can fully pay off.

Reframing AES as a cost-constrained measurement system helps supply that advance. It places cost inside the logic of assessment design, models compute burden as a multi-layer construct, and insists that comparisons be made under quality-cost sufficiency rather than under raw performance rankings. External literatures matter because they show that the necessary accounting logic is already mature, while emerging AES evidence shows that the pressure is no longer hypothetical. The CAMD Framework is therefore offered not as a way to rank models, benchmark vendors, or estimate a single cloud bill. It gives testing programs and vendors a common vocabulary for metering and reporting the cost of scoring; it gives researchers a reporting standard that makes cost claims comparable across studies and hardware environments; and it gives score users a basis for asking whether cost-saving choices have compromised validity, reliability, or fairness.